\documentclass[aps,twocolumn,amsmath,amssymb,showpacs,prb,unsortedaddress]{revtex4-1} 
\usepackage{epsf}   
\usepackage{graphicx}
\usepackage{gensymb}
\usepackage{float}

\begin{document}

\title{Experimental and theoretical electronic structure of EuRh$_2$As$_2$}

\author{A. D. Palczewski} 
\author{R. S. Dhaka}
\author{Y. Lee}
\author{Yogesh Singh}
\altaffiliation{Present address: Indian Institute of Science Education and Research Mohali, Chandigarh 140306, India.}
\author{D. C. Johnston}
\author{B. N. Harmon}
\author{Adam Kaminski}
\email {kaminski@ameslab.gov} 

\affiliation{Division of Materials Science and Engineering, 
The Ames Laboratory, US DOE, and Department of Physics and Astronomy, Iowa State University, Ames, Iowa 50011, USA}

\date{\today}

\begin{abstract}
The Fermi surfaces (FS's) and band dispersions of EuRh$_2$As$_2$ have been investigated using angle-resolved photoemission spectroscopy. The results in the high-temperature paramagnetic state are in good agreement with the full potential linearized augmented plane wave calculations, especially in the context of the shape of the two-dimensional FS's and band dispersion around the $\Gamma$ (0,0) and ${\rm X}$ $(\pi,\pi)$ points.  Interesting changes in band folding are predicted by the theoretical calculations below the magnetic transition temperature $T_{\rm N}\approx 47$~K. However, by comparing the FS's measured at 60~K and 40~K, we did not observe any signature of this transition at the Fermi energy indicating a very weak coupling of the electrons to the ordered magnetic moments or strong fluctuations. Furthermore, the FS does not change across the temperature ($\approx 25$~K) where changes are observed in the Hall coefficient. Notably, the Fermi surface deviates drastically from the usual FS of the superconducting iron-based {\it A}Fe$_2$As$_2$ parent compounds, including the absence of nesting between the $\Gamma$ and X FS pockets. 
\end{abstract}


\maketitle

The recent discoveries of superconductivity in FeAs-based materials, {\it R}LaFeAs(O$_{1-x}$F$_x$) where {\it R} is a lanthanide element\cite{Kamihara08,ChenNature08,Chen08,ChenPRL08,Takahashi08} and (Ba$_{1-x}$K$_x$)Fe$_2$As$_2$,\cite{Rotter08,Yuan09} resulted in a large number of experimental and theoretical studies. These materials have revealed fascinating properties regarding the complex interplay among structure, magnetism and superconductivity,\cite{Cruz08} especially the {\it A}Fe$_2$As$_2$ ({\it A} = Ba, Sr, Ca, Eu) family of compounds with $T_{\rm c, max}\approx 38$~K.\cite{SefatPRL08,Ni08,Sasmal08,Kreyssig08,Goldman09,Kurita11,Jeevan11} It has been shown that the parent compounds manifest simultaneous transitions where the high-temperature tetragonal paramagnetic phase changes at $\approx 140-205$~K to the lower-temperature orthorhombic phase with antiferromagnetic order which is associated with a spin density wave (SDW) and superconductivity is achieved in different ways.\cite{Rotter08,SefatPRL08,Ni08,CanfieldRev10,Colombier09,Kimber09,Johnston10} There has been a flurry of activity trying to understand the basic properties of these new materials, and in particular, the mechanism for high $T_{\rm c}$ where the Fermi surfaces (FS's) play an important role.\cite{Liu,Fink09,Jensen11}  

The key feature common to these materials is the presence of stacked FeAs layers. This gives a strong motivation to investigate similarly structured compounds in a search for additional high $T_{\rm c}$ superconductors and to understand the mechanism of the superconductivity and magnetic ordering. More recently, several isostructural materials have been found which show very interesting physical properties and can be potential parent compounds for high $T_{\rm c}$ superconductors, for example, BaMn$_2$As$_2$,\cite{DavidMn}, EuRu$_2$As$_2$,\cite{Jiao11} EuRh$_2$As$_2$,\cite{Nandi09, SinghPRBR09} BaRh$_2$As$_2$,\cite{Singh09} and SrRu$_2$As$_2$.\cite{Nath09} The electronic structure of BaNi$_2$As$_2$, in particular, shows no signature of band folding suggesting the absence of SDW magnetic ordering.\cite{Zhou11} Because these materials are not superconducting, the absence of Fe in these materials allows us to determine what role the Fe atoms play in superconductivity. Particularly, EuRh$_2$As$_2$ shows unusual characteristics that are not observed in the superconducting materials, including giant magnetoresistance and a strong reduction in the electronic specific heat coefficient with applied field in the antiferromagnetic state.\cite{SinghPRBR09}  Magnetic scattering measurements reveal that the Eu spins are ferromagnetically aligned within the $a$-$b$ plane where the spins between adjacent Eu planes are nearly antiparallel.\cite{Nandi09} A previous calculation suggested that the maximum contribution in the electronic density of states at the Fermi energy ($E_{\rm F}$) is from the Rh 4$d$ orbitals.\cite{SinghPRBR09} Considering that the electronic states near the FS are dominated by contributions from the transition metal element, understanding the interplay between Rh and As at the FS is vital for these materials. To the best of our knowledge, there are no previous reports on the Fermi surfaces of EuRh$_2$As$_2$. 

Here, we present the first angle-resolved photoemission spectroscopy (ARPES) study on EuRh$_2$As$_2$ detailing the three dimensional nature of the FS and comparing it to the theoretical full potential-linearized augmented plane-wave (FP-LAPW) calculations. We find that the FP-LAPW calculations predict the shape of the FS and general band dispersion quite well especially when comparing the data in the paramagnetic phase. Magnetic susceptibility and heat capacity measurements have demonstrated that EuRh$_2$As$_2$ undergoes a transition from a paramagnetic state to an antiferromagnetic state below the N\'eel temperature $T_{\rm N}\approx 47$~K.\cite{SinghPRBR09} The ARPES data in the proximity of $E_{\rm F}$ do not show any signatures of this transition. Interestingly, the FS and band structure of EuRh$_2$As$_2$ are very different from those of other similar compounds including EuFe$_2$As$_2$\cite{Zhou10,Jong10} where the superconducting family has a hole band centered at $\Gamma$ and electron band centered at ${\rm X}$ which are closely nested.\cite{Liu,Thirupathaiah11} On the other hand, EuRh$_2$As$_2$ has electron bands centered at $\Gamma$ and ${\rm X}$ with no evidence of nesting. 

\begin{figure*}
\includegraphics[width=6.95in]{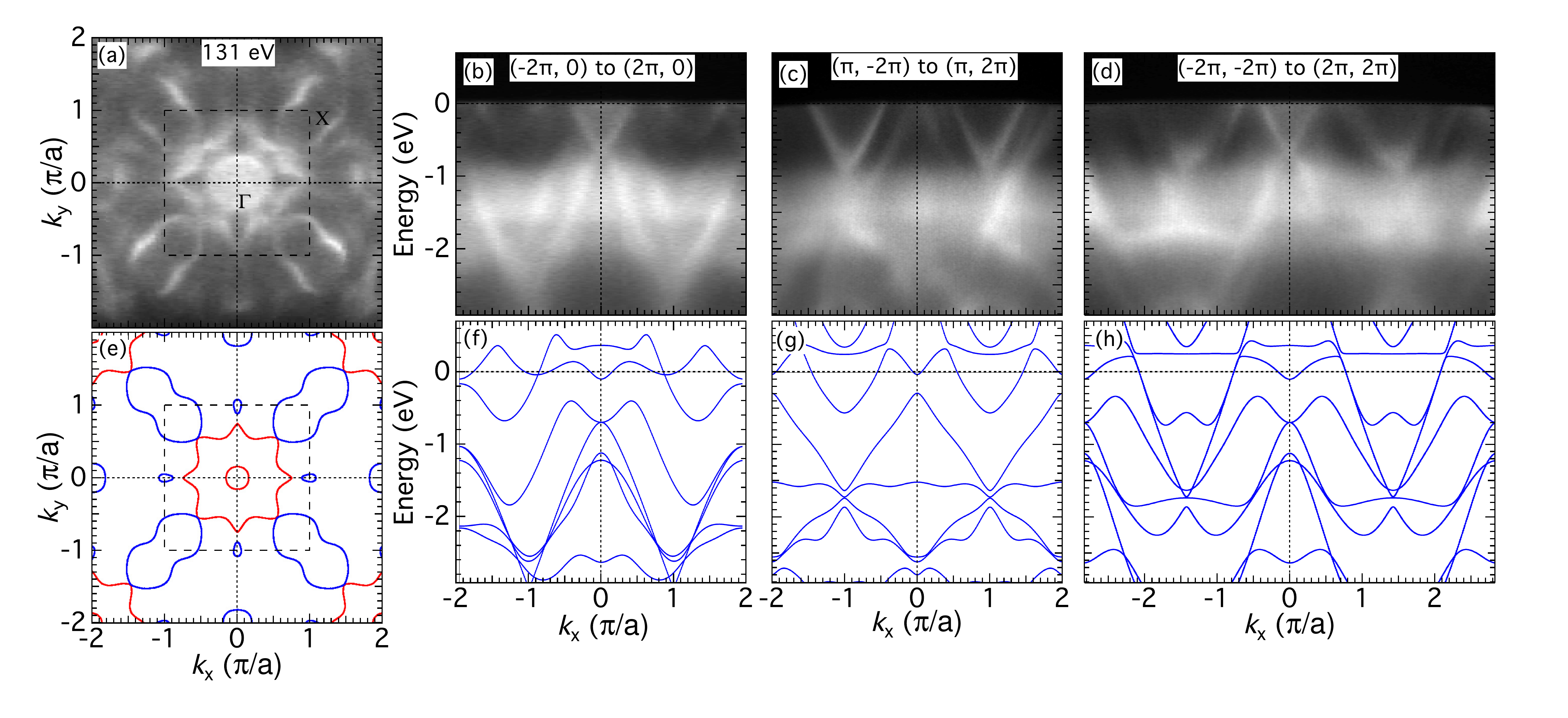}
\begin{centering}
\caption{\label{Figure1}(Color online) Comparison between the measured ARPES data (40~K) and the theoretical FP-LAPW calculations in the paramagnetic (high temperature) state: (a) measured FS of EuRh$_2$As$_2$ plotted within 50~meV of $E_{\rm F}$ over the first two Brillouin zones (BZ's) at 131~eV; (b--d) band dispersions from (a); (e--h) the same as (a--d) but calculated. The dashed squares in (a) and (e) bounded by $-1$ to $1$ in $k_{ x}$ and $k_{ y}$ mark the first BZ boundary of the paramagnetic state (body-centered tetragonal structure). The directions of the band cuts are shown in each panel. The FP-LAPW FS calculation at $k_z=3.5~\pi/c$ (h$\nu=131$~eV) in (e) shows the predominant contributions from the Rh 4$d$ (red) and the As 4$p$ (blue) bands which cross $E_{\rm F}$.}
\end{centering} 
\end{figure*}

Single crystals of EuRh$_2$As$_2$ were grown out of Pb flux. X-ray diffraction measurements confirmed that EuRh$_2$As$_2$ crystallizes in the tetragonal ThCr$_2$Si$_2$ structure with lattice parameters $a=4.075(4) $~\AA\ and $c=11.295(2) $~\AA\ at 298~K.\cite{SinghPRBR09} 
Single crystal samples were cleaved {\it in situ} at the base pressure of $\approx4\times10^{-11}$~mbar. ARPES measurements were performed by using a Scienta R4000 electron analyzer on beamlines 7.0.1 and 10.0.1 at the Advanced Light Source (ALS), Berkeley, California. The energy and momentum resolution were set to $\sim20$~meV and $\sim0.3^{\circ}$, respectively.   

The FP-LAPW method with the local density approximation\cite{Perdew92} was used to calculate the theoretical FS and band dispersions. To obtain self-consistent charge density, we employed $R_{\rm MT}\times{\it k}_{\rm max}=8.0$  (the smallest muffin tin radius multiplied by the maximum $k$ value in the plane wave expansion basis) with muffin tin (MT) radii of 2.5, 2.2, and 2.2~$a_{\rm 0}$ (Bohr radius) for Eu, Rh, and As, respectively.  The calculations were carried out at 475 $k$-points in the irreducible Brillouin zone and the calculations were iterated to reach a total energy convergence criterion of  0.01 mRy/cell. While in the paramagnetic calculations the 4$f$ electrons were treated as core electrons,  a local-density-approximation plus Coulomb potential (LDA+U) method with $U=5$~eV was used for the 4$f$ electrons in the antiferromagnetic calculations. 

For the FS measurements, two different photon energies of 131~eV (Figs.~1 and 3) and 105~eV (Fig.~2) were chosen by taking a map at $k_{ x,y}(\pi/a)=0$ with varying incident photon energy (along the $k_{||}$--$k_z$  plane), {\it i.e.} $k_{z}$ dispersion (not shown). Note that, in the photoemission process, the component of the momentum of the outgoing electron perpendicular to the surface, {\it i.e.} $k_{z}$, is not conserved and creates an offset with respect to the high symmetry points. This nonconservation is due to the surface potential inherent in all materials, which can be calculated by measuring a $k_{ z}$ dispersion map and using $k_{ z}=(1/{\hbar})\sqrt {2m(E_{\rm k}{\rm cos}^{2}\theta+V_{0})}$, where $E_{\rm k}$ is the photoelectron kinetic energy and $V_0$ is the inner potential. In the map, we observe a non-integer $k_{z}$ value at the symmetry points as compared to the calculations. Therefore, we included an energy shift (inner-potential) into the $k$ value calculations which aligns the symmetry points to an integer $k_{z}$ value at the symmetry point.\cite{Hufner95,Starowicz08}  In the present case, the offset or inner-potential is estimated to be 8.9~eV by setting the $k_{z}=18~\pi/c$ value of symmetry point $\Gamma$ to be  0~$\pi/c$. The two-dimensional FS plots of EuRh$_2$As$_2$ are shown in Figs.~1(a) and 2(a) measured using photon energies h$\nu=$131~eV ($k_{z}=3.5~\pi/c$) and 105~eV ($k_{z}=1.15~\pi/c$), respectively, which correspond to two different symmetry points. It is quite remarkable that the FS [Figs.~1(a) and 2(a)] of EuRh$_2$As$_2$ are very different from those of {\it A}Fe$_2$As$_2$ (where {\it A} = Ba, Sr or Ca\cite{Kondo10} as well as Eu\cite{Zhou10,Jong10}), where both electron and hole pockets are nearly nested. 

  \begin{figure*}
\includegraphics[width=6.9in]{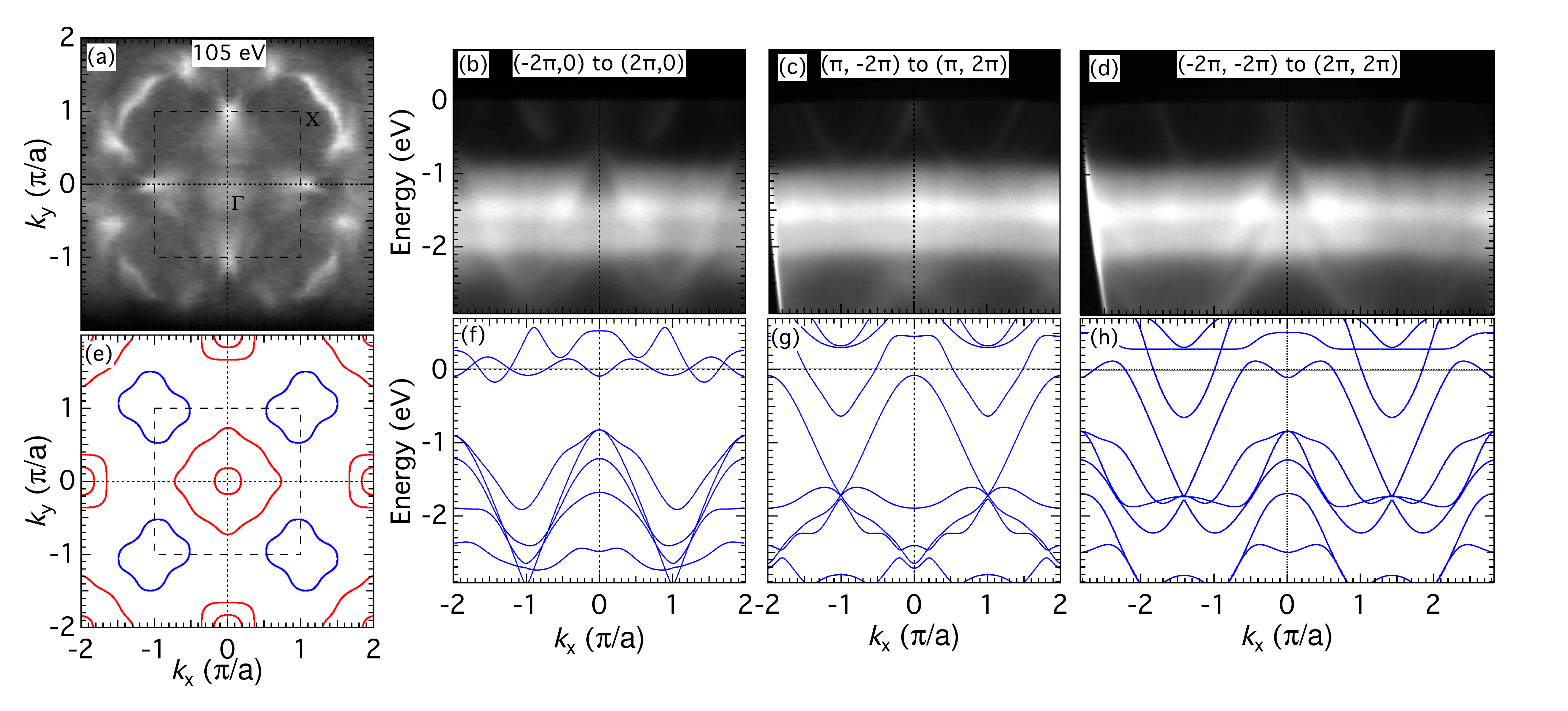}
\begin{centering}
\caption{\label{Figure2}(Color online) The FS and band dispersion plots, as defined in the caption of Fig.~1, except measured and calculated at $k_{z}=1.15~\pi/c$ (h$\nu=105$~eV). }
\end{centering}
\end{figure*}  

Comparison of the measured ARPES data with theoretical calculations is vital for understanding the electronic structure of these materials. For this purpose, we have shown the experimental FS map at 131~eV ($k_{\rm z}=3.5~\pi/c$) in the paramagnetic phase at 40K and the corresponding FP-LAPW calculations around the $\Gamma$ $(0,0)$ point in Figs.~1(a) and 1(e), respectively. Both the measured as well as calculated FS's show two pockets centered at the $\Gamma$ point and another pocket centered at ${\rm X}$ $(\pi,\pi)$. This agreement shows that the FP-LAPW calculations reproduce our ARPES data quite well. Although the general shape of the pockets match, the size of these pockets do not, where the pocket at $\Gamma$ is larger for the measured one. Unlike the FS at 105~eV (Figs.~2(a) and 2(e)), the FS pockets at the ${\rm X}$ point in Figs.~1(a) and 1(e) are quite similar without a significant matrix element effect. To examine the character of the FS pockets at the $\Gamma$ and ${\rm X}$ points as measured using ARPES, which can usually be easily determined by tracing the dispersions of the associated bands, we compared the band dispersions along the symmetry points. The measured band dispersions are shown in Figs.~1(b--d) and the calculated band cuts are shown in Figs.~1(f--h). The symmetry cut lines are mentioned in the respective plots in Figs.~1 and 2. It is clear that the bands observed at the $\Gamma$ and ${\rm X}$ points show an electron-like nature. All the measured band dispersions in Figs.~1(b--d) are in agreement with the respective calculations in Figs.~1(f--h) in terms of the shape. However, the depths of the bands are shallower in the calculated data which cause a change in the sizes of the FS pockets.  

\begin{figure}
\includegraphics[width=3.0in]{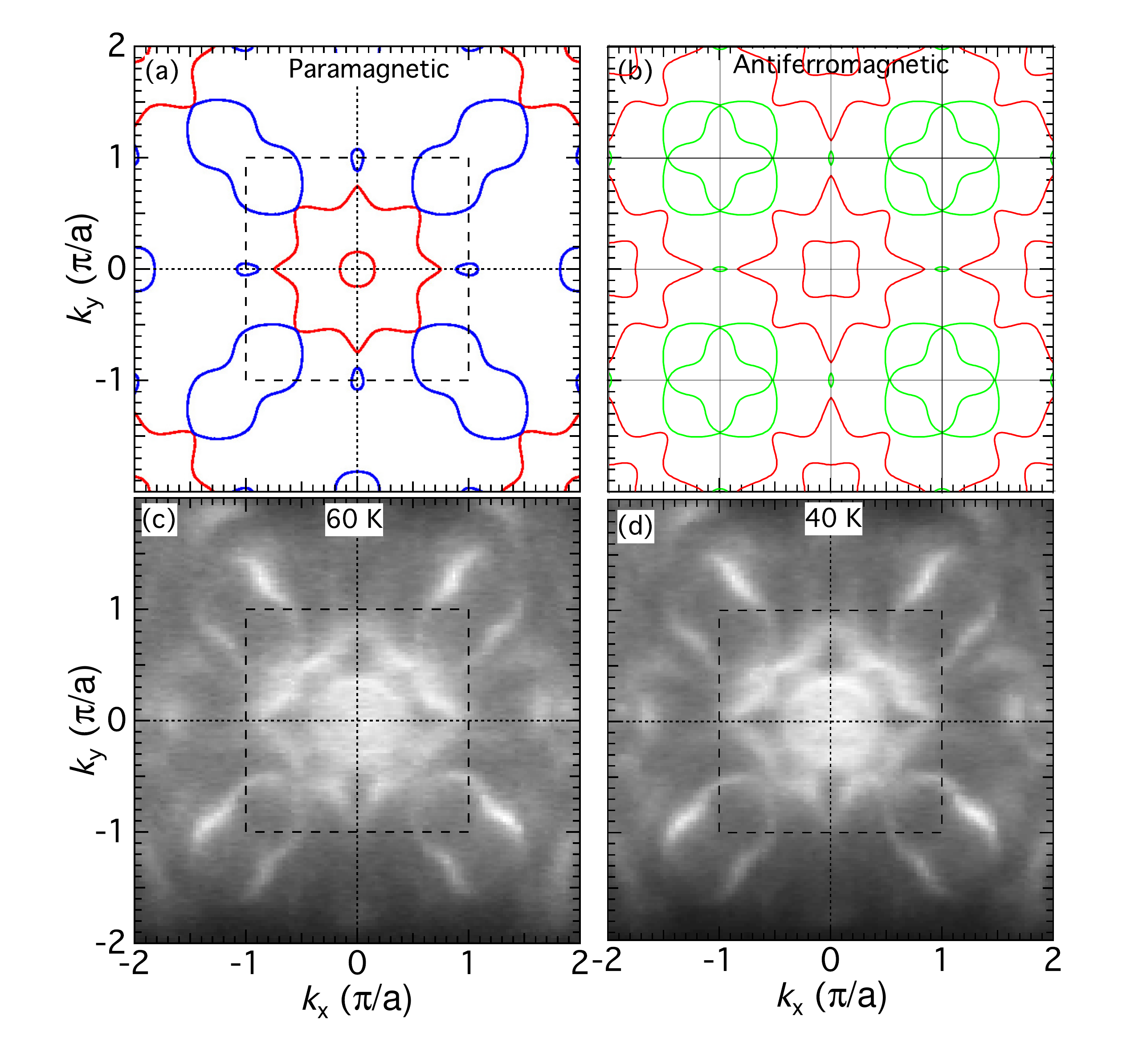}
\begin{centering} 
\caption{\label{Figure3}(Color online) Comparison of the FP-LAPW calculations and the FS measured at $k_{z}=3.5~\pi/c$ (h$\nu=$131~eV) by ARPES above and below $T_{\rm N}\approx47$~K. Theoretically calculated FS at (a) high temperature paramagnetic and (b) low temperature antiferromagnetic states showing band back-folding caused by a doubling of the unit cell when the sample goes from a nonmagnetic body-centered tetragonal to a magnetic tetragonal structure. Colors are used for clarity and do not reflect particular orbital contributions. The measured FS at (c) 60~K paramagnetic and (d) 40~K antiferromagnetic states showing no change in the band structure.}
\end{centering} 
\end{figure}

To understand the three-dimensional nature of EuRh$_2$As$_2$ in detail, another photon energy h$\nu=105$~eV ($k_{z}=1.15~\pi/c$) was chosen where the FS was measured and calculated [Figs.~2(a) and 2(e), respectively] along the $a$-$b$ plane, which also show two pockets centered at $\Gamma$ and another pocket centered at the ${\rm X}$ points (Fig.~2). The sizes of the $\Gamma$ pockets are larger in the measured data than in the calculated data, as similarly observed for h$\nu=131$~eV (Fig.~1). The strong crossing of the pocket centered at ${\rm X}$ in the second BZ matches the shape of the calculated data, yet in the first BZ, the bands are mostly absent. This absence could be from an ARPES matrix element effect where the band transitions are not allowed and therefore cannot be observed.\cite{Damasceli03}  By comparing the measured band dispersions in Figs.~2(b--d) with the calculated ones in Figs.~2(f--h), the nature of these pockets is clearly electron-like. Note that the FS of the {\it A}Fe$_2$As$_2$-type superconducting pnictides consist of closely nested pockets at $\Gamma$ and ${\rm X}$,\cite{Liu,Liu11,Yi09,Terashima09,Liu09,Yang09} where the contributions from the Fe $3d$ orbitals dominate. This is not the case for EuRh$_2$As$_2$ where the FS pockets are electron-like with the contributions mainly from both Rh 4$d$ electron bands at the $\Gamma$ point and As 4$p$ electron bands at the ${\rm X}$ point. 

\begin{figure}
\includegraphics[width=3.30in]{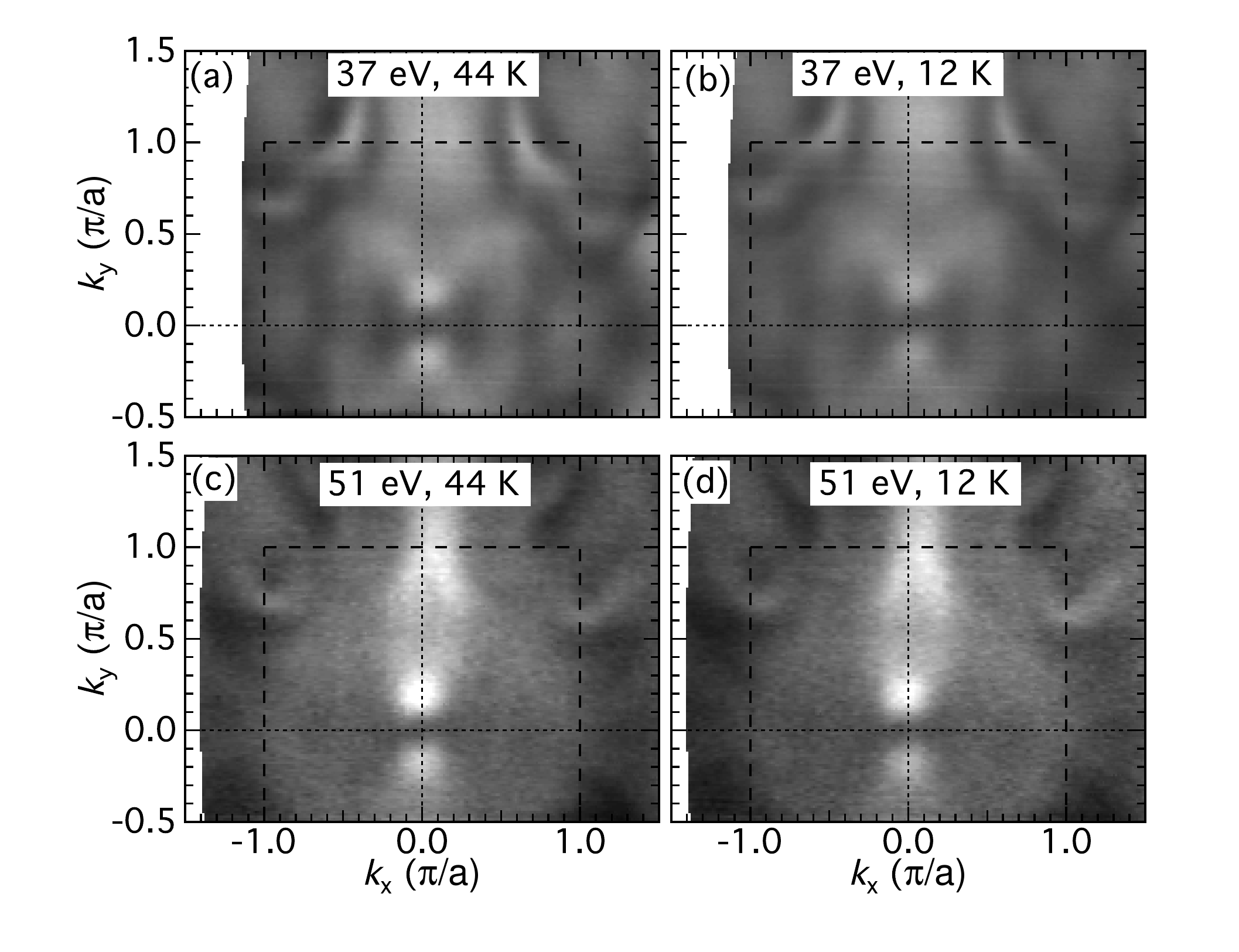}
\begin{centering} 
\caption{\label{Figure4} The FS of EuRh$_2$As$_2$ above (44K) and below (12K) the temperature ($\sim25$~K) where the Hall coefficient changes sign, plotted within $\pm$ 20~meV of $E_{\rm F}$. The photon energy and measured sample temperature are shown in each panel.}
\end{centering}
\end{figure}

Previous transport studies on EuRh$_2$As$_2$\cite{SinghPRBR09} revealed two transition or crossover temperatures. These are the N\'eel  temperature  $T_{\rm N}\approx47$~K and the temperature where the Hall coefficient changes sign at $T\sim25$~K.  Above $T_{\rm N}$,  EuRh$_2$As$_2$ is in the paramagnetic body-centered-tetragonal state.  Below $T_{\rm N}$, the Eu moments are ferromagnetically aligned within the tetragonal  $a$-$b$ plane and nearly antiferromagnetic along the $c$ axis.\cite{Nandi09} The theoretical FP-LAPW calculations show a change in the FS upon crossing $T_{\rm N}$, shown in Figs.~3(a) and 3(b) in the paramagnetic and antiferromagnetic phase, respectively. The calculations show a reduction of the BZ below $T_{\rm N}$ because the magnetic phase transition breaks the body-centered symmetry and doubles the unit cell. However, the ARPES data [Fig.~3(c) at 60~K and 3(d) at 40~K] do not show any signature of this back-folding. This might arise if the electrons do not couple to the ordered magnetic moments of the Eu ions or presence of significant fluctuations.\cite{Jong10} The Eu is divalent with seven 4f electrons.  The Eu 5d bands begin about 1eV above the Fermi level, resulting in weak RKKY coupling and low ordering temperature.Another possibility is that the AF propagation vector in the $k_{z}$ direction could cause the bands to overlap and hence no change would be present in the FS measured in the $a$-$b$ plane. A previous study on EuRh$_2$As$_2$ showed no significant change in resistivity across $T_{\rm N}$, while sharp transitions were observed at $T_{\rm N}$ in heat capacity and susceptibility measurements.\cite{Singh09} For EuFe$_2$As$_2$, it was shown that the Eu magnetic ordering does not have significant influence on the pre-existing Fe SDW ordering.\cite{Jong10}

Another transition or crossover occurs in EuRh$_2$As$_2$ where the Hall coefficient changes sign from negative to positive at around 25~K,\cite{SinghPRBR09} which is well below the magnetic transition temperature. This sign change was suggested to be a possible signature of temperature induced carrier redistribution between electron and hole like Fermi surfaces. In order to ascertain how the sign change in Hall coefficient would effect the FS, we measured the FS above and below this temperature (Fig.~4) {\it i.e.}~at 44~K and 12~K, respectively. In Figs.~4(a) and 4(b), we present the FS measured at 37~eV, while Figs.~4(c) and 4(d) show the FS measured at 51~eV\@. Surprisingly, the data show no change in the FS between 44~K and 12~K. Although unlikely, changes might occur in the FS at other than used here photon energies. 

In conclusion, the Fermi surface of EuRh$_2$As$_2$ has been studied using ARPES and compared with the theoretical FP-LAPW calculations. The FP-LAPW calculations map the general shape of the FS and band dispersion quite well especially when compared to the high temperature paramagnetic data. We observed the signature of the three-dimensional nature of the FS. Surprisingly, the FS data do not show any indication of the AF state below the $T_{\rm N}$ implying a weak coupling between the Eu and RhAs layers. Moreover, the sign change in the Hall coefficient below 25~K is not visible in the FS maps. Notably, the band structure of EuRh$_2$As$_2$ is very different than {\it A}Fe$_2$As$_2$ compounds including EuFe$_2$As$_2$\cite{Zhou10,Jong10} where the superconducting family has nested hole and electron bands centered at $\Gamma$ and ${\rm X}$, respectively,\cite{Liu11} whereas EuRh$_2$As$_2$ has electron bands centered at $\Gamma$ and ${\rm X}$ without nesting. In this respect, the electronic structure presented here is strikingly different. 

We thank Eli Rotenberg and Sung-Kwan Mo for excellent support at the ALS. The work at the Ames Laboratory was supported by the Department of Energy-Basic Energy Sciences under Contract No. DE-AC02-07CH11358. The ALS is supported by the U.S. DOE under Contract No. DE-AC02-05CH11231.

\end{document}